\documentstyle[11pt,paspconf,epsfig]{article}

\begin{document}

\title{The formation epoch of massive ellipticals, as traced by radio galaxies}

\author{J. S. Dunlop}
\affil{Institute for Astronomy, University of Edinburgh, Royal Observatory, 
Edinburgh, EH9 3HJ}

\begin{abstract}
I review the current status of an on-going investigation into the stellar 
and dynamical ages of the oldest known galaxies at $z > 1$. The
spectroscopic data, when compared with the predictions of a range of 
recent evolutionary synthesis models, continue to indicate that the
oldest galaxies at $z \simeq 1.5$ are already $> 3$ Gyr old. Moreover,
comparison with new models 
which incorporate chemical evolution, suggests 
that this result may be less sensitive to the assumed metallicity than
was previously suspected. Such ages should therefore be taken seriously
as constraints both on theories of galaxy formation and on cosmological 
models. Dynamically these objects also appear well
evolved; they display de Vaucouleurs surface brightness profiles, and 
I demonstrate 
that their HST-derived morphological parameters place them on the same 
Kormendy relation as the $z \simeq 1$ 3CR galaxies. Finally I consider 
the ages of these intermediate redshift objects in the context of our
recently completed SCUBA
sub-mm survey of radio galaxies spanning the redshift range 
$z = 1 \rightarrow 4$. These
data indicate that the main epoch of star-formation in radio galaxies
lies at $ z \simeq 4$, a result which fits naturally
with the ages of the oldest galaxies at $ z = 1.5$ within an open Universe.
\end{abstract}


\section{Introduction}

Even a very low-level burst of star-formation activity can easily obscure the 
true colours of the dynamically-dominant stellar population in a high-redshift
galaxy. For this reason it is necessary to use the reddest (unreddened) 
galaxies at a given redshift to set meaningful constraints on their 
primary formation epoch. As demonstrated by Dunlop et al. (1996), 
Spinrad et al. (1997) and Dunlop (1999) the reddest ($R-K \simeq 6$) optical 
identifications of milli-Jansky radio sources have provided the best 
examples discovered to date of passively evolving galaxies at redshifts 
$z > 1$. As discussed by Dunlop (1999) there can be little doubt that the
red colours of 53W091 ($R-K = 5.8$; $z = 1.55$) and 53W069 ($R-K = 6.3$;
$z = 1.43$) are due to evolved stellar populations rather than dust.
However, the initial attempt at accurate age-dating of 53W091 by Dunlop
et al. (1996) has since stimulated considerable debate and controversy 
(Bruzual \& Magris 1997, Yi et al. 1999, Heap et al. 1998).

In the next section I summarize the {\it uncontroversial} facts about these
objects. I then consider the controversy over their precise ages and, 
focussing on 53W069, demonstrate that there is in fact rather
{\it little disgreement} between the ages inferred from different
evolutionary synthesis models, provided fitting is confined 
to the high-quality rest-frame near-UV spectra. 
Following this I revisit the issue of
age-metallicity degeneracy before moving on to consider the morphologies
of these galaxies, and their ages in the context of our recently completed
major SCUBA survey of dust-enshrouded star-formation in high-redshift
radio galaxies.

\section{53W091 and 53W069: uncontroversial facts}

The Keck spectra of 53W091 (Dunlop et al. 1996; Spinrad et al. 1997) and
53W069 (Dunlop 1999; Dey et al. in preparation) 
both display strong spectral breaks at
rest-frame wavelengths of 2640\AA\ and 2900\AA\ (along with several
other repeatable absorption features) which prove beyond doubt that
their ultraviolet spectra are dominated by starlight from F stars. This
is illustrated in Figure 1; the near-ultraviolet SED of 53W091 is essentially
indistinguishable from an F5V star, while that of 53W069 is best described
by an even redder F9V star. Since for ages less than $\simeq 5$ Gyr essentially
all spectral synthesis models predict that the {\it near-ultraviolet}
light of a stellar population should be dominated by main-sequence stars,
this means that the main-sequence turnoff point in these stellar
populations must have evolved into the mid F-star regime. 

Figure 1 also
shows that 53W069 appears to be a cleaner example of a genuinely coeval
stellar population; the SED of 53W091 can be constructed by adding a
low-level population of F0V stars (or some other
comparably blue low-level component) to the SED of 53W069.
Deriving the ages of these objects thus boils down to `simply' 
determining how long it takes for the near-ultraviolet light from an
evolving main sequence to impersonate the SED of an F9V star.
It is worth noting that, at the time of writing, these two galaxies
remain the only objects at redshifts as high as $z \simeq 1.5$ for which
this type of potentially accurate age dating is possible.

\begin{figure}
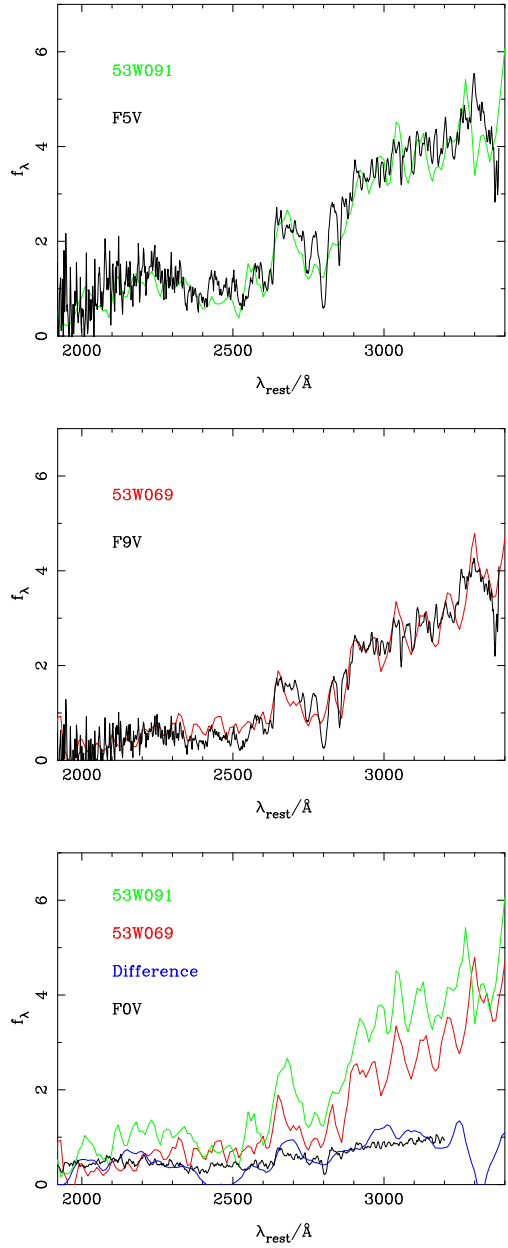

\vspace{38pc}
\includegraphics{dunlopj1a.eps}
\includegraphics{dunlopj1b.eps}
\includegraphics{dunlopj1c.eps}
\caption{\small Top Panel: Galaxy rest-frame Keck spectrum of
53W091 compared with an average F5V IUE spectrum.
Middle Panel: Galaxy rest-frame Keck spectrum of 53W069 compared with an
average F9V IUE spectrum. Bottom Panel: Comparison of the properly scaled 
rest-frame UV spectra of 53W091 and 53W069 with a smoothed version of the 
difference spectrum compared with an average F0V IUE spectrum.}
\end{figure}

\section{Age controversy}

\subsection{Model dependence}

Following the discovery of 53W091, Dunlop et al. (1996) derived an 
estimate for its age of 3.5$\pm 0.5$ Gyr. This estimate was 
based primarily on comparison with a 
main-sequence only model of spectral evolution, with special
emphasis placed on the (reddening independent) strength of the the 2640\AA\ 
and 2900\AA\ spectral breaks. The age-dating of this galaxy was then
explored in more detail by Spinrad et al. (1997) who highlighted the
disagreement between different models (especially if $R-K$ colour was
included as a fitted quantity), but again concluded that 
an age of 3.5 Gyr appeared to be the most reasonable estimate of the time
elapsed since cessation of star-formation activity.
Subsequently, however, both Bruzual \& Magris (1997) and Yi et al. (1999)
have concluded in favour of an age as young as 1.5 Gyr.

However, it transpires that 
a significant fraction of this disagreement arises from the fact
that Bruzual \& Magris (1997) and Yi et al. (1999) have included $R-K$
colour as an important factor in the fitting process. This dilutes much
of the advantage offered by the detailed rest-frame near-ultraviolet
Keck spectra obtained for these objects, because the evolution of $R-K$
colour at such redshifts depends not only on the main-sequence, but
also on modelling of the later stages of stellar evolution, over which
there remains much more controversy (due to complications such as mass
loss; see Jimenez et al. 1999). In fact, if fitting is confined to the
Keck spectrum, the models of Bruzual and of 
Yi indicate an age of $\simeq$2.5 Gyr,
while the most recent models of Jimenez indicate an age of 3.0 Gyr.
Given the uncertainties, these ages are basically consistent, albeit 0.5
- 1 Gyr younger than the original age quoted by Dunlop et al. (1996)
(although the spectral breaks, especially the 2900\AA\ break continue to favour an
older age $\simeq 4$ Gyr - see Dunlop 1999).

\begin{figure}
\vspace{30pc}
\includegraphics{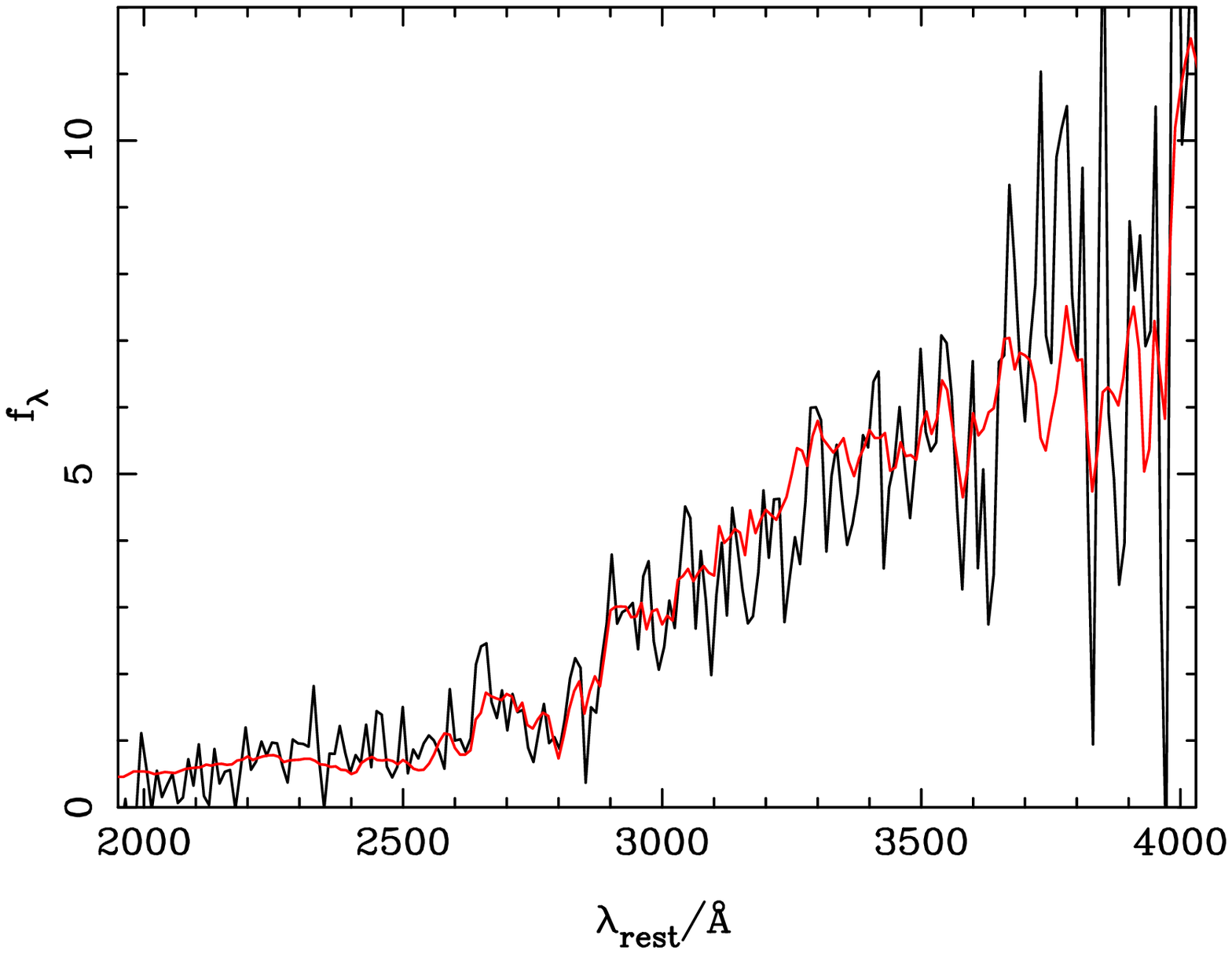}
\includegraphics{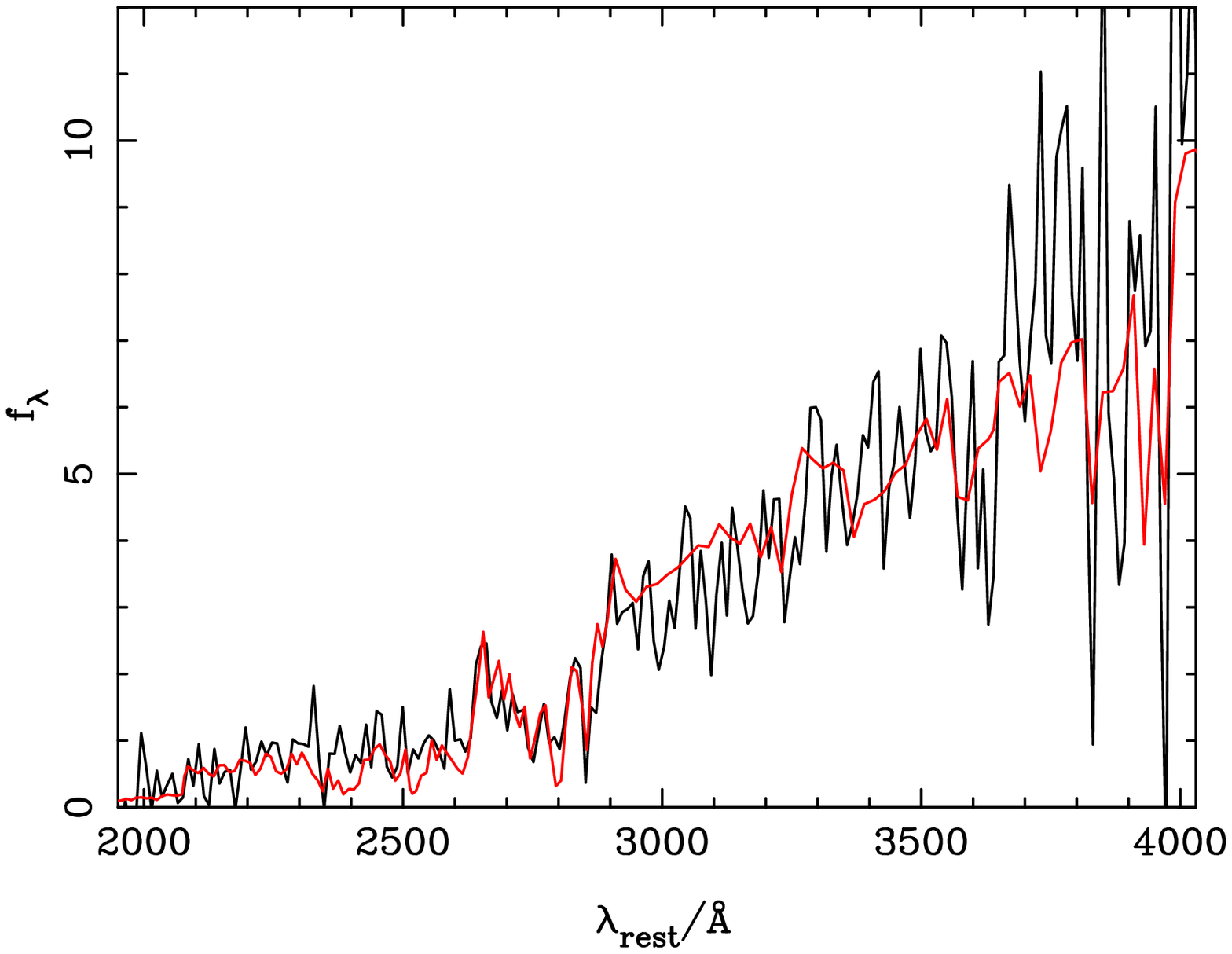}
\caption{\small Upper panel: The spectrum of 53W069 overlaid with the
best fitting Bruzual-Charlot model, which has an age of 3.25 Gyr.
Lower panel: The spectrum of 53W069 overlaid with the best fitting model
from Jimenez et al. (1999), which has an age of 4.25 Gyr.}
\end{figure}

\begin{figure}
\vspace{17pc}
\includegraphics{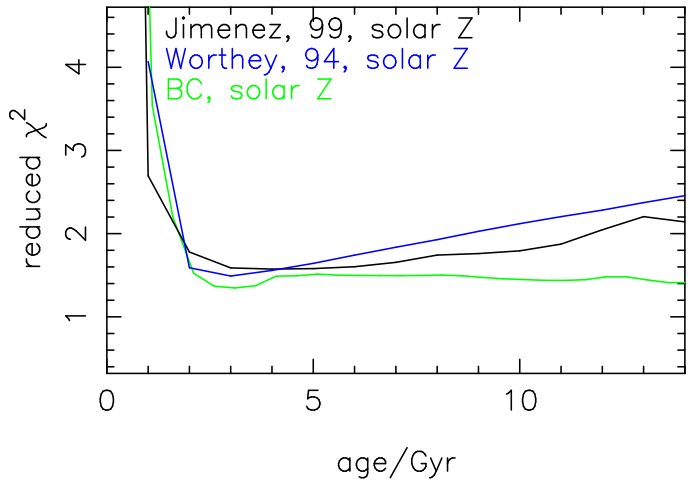}
\caption{\small Reduced chi-squared versus the inferred age of 53W069
derived from attempting to fit the Keck spectrum of 53W069 with the
evolutionary synthesis models
of Jimenez et al. (1999), Bruzual \& Charlot (1993) and Worthey (1994).}
\end{figure}

However, as explained above, 53W069 provides a better example of a
clean coeval stellar population, and for this galaxy it appears that
essentially all existing models indicate an age $ > 3$ Gyr. Figure 2
shows the data fitted by the updated models of Bruzual \& Charlot (1993) (inferred age
3.25 Gyr) and by the most recent models of Jimenez et al. (1999) 
(inferred age 4.25 Gyr).

\subsection{Age constraints for 53W069}

Figure 3 shows a plot of reduced chi-squared versus inferred age,
derived from attempting to fit the models of Jimenez et al. (1999), 
Bruzual \& Charlot,
and Worthey (1994) to the Keck spectrum of 53W069. 
The models of Jimenez favour an age about 1 Gyr older than the
models of Bruzual \& Charlot, but it is clear from this plot that this
disagreement is not very dramatic in terms of quality of fit.
The key point is that all the models favour
an age of 3 Gyr or greater, with ages younger than 2.5 Gyr formally
excluded.
It is important to re-stress that these ages are based on
instantaneous starburst models, and thus indicate the time elapsed since
cessation of major star-formation activity with no additional time included
for the process of galaxy/star formation. However, they do assume solar 
metallicity.

\subsection{Impact of varying metallicity}

\begin{figure}
\vspace{17 pc}
\includegraphics{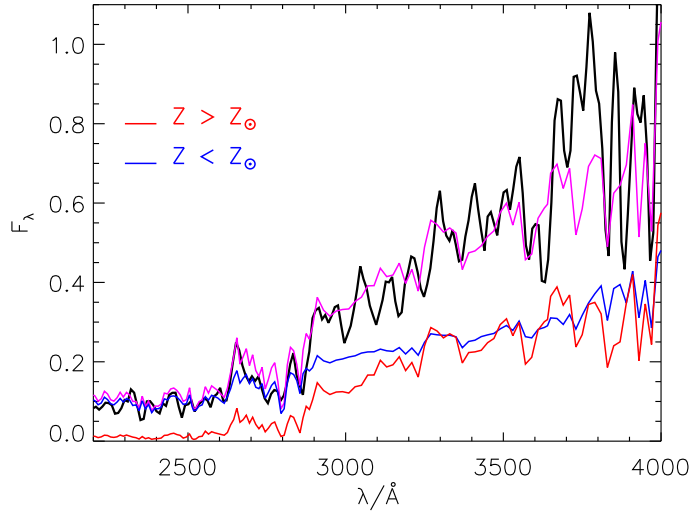}
\caption{\small The Keck spectrum of 53W069 fitted with the composite 
metallicity model of Jimenez et al. (1999) at an age of 4 Gyr. The
mass-averaged metallicity of this model is $\simeq$ 2 $\times$ solar, but
the figure also illustrates how the sub-solar metallicity stars continue
to dominate the predicted SED shortward of 3500\AA, leading to an
age estimate which is unexpectedly robust to varying metallicity.}
\end{figure}

A major concern with this type of spectral age dating is that, as
explored by for example Worthey (1994), 
deduced age is approximately inversely
proportional to assumed metallicity for most spectral age indicators.
For the 2640\AA\ and 2900\AA\ breaks this age/metallicity degeneracy
appears to be not quite as severe as this (Dunlop et al. 1996; Fanelli et al.
1992), but nonetheless it is clear that the inferred ages of 53W091 and 53W069 
can be reduced to less than 2 Gyr if one assumes the observed
stellar population consists only of stars with twice solar metallicity.
 
However, new models of elliptical galaxy formation/evolution
developed by Jimenez et al. (1999), and by Yi et al. (1999) suggest that 
adoption of this simple-minded age-metallicity degeneracy may lead to an
overly pessimistic view of the accuracy to which ages can be derived.
These models include chemical evolution during the initial starburst, in
an attempt to produce a realistic stellar population of composite 
metallicity. The interesting consequence of
comparing the predictions of such models to the near-ultraviolet 
SED of 53W069 is 
that the inferred age is essentially unchanged from that derived using the 
simple solar metallicity models, despite the fact that the mass-weighted
metallicity can be as high as twice solar. The reason for this is
illustrated in Figure 4 which shows the best fit obtained using the
$2 \times$-solar metallicity composite model of Jimenez et al. (1999b) to
53W069, at an age of 4 Gyr. The plot shows that while the high metallicity
stars dominate the red end of the spectrum, sub-solar metallicity stars
continue to dominate shortward of $\lambda \simeq 3300$\AA, with the
consequence that the derived age is more robust than would have been
niavely expected. Indeed the statistical 
fit obtained with this 4 Gyr composite model
is substantially better than that achieved with any single metallicity
model.

\begin{figure}
\vspace{18pc}
\includegraphics{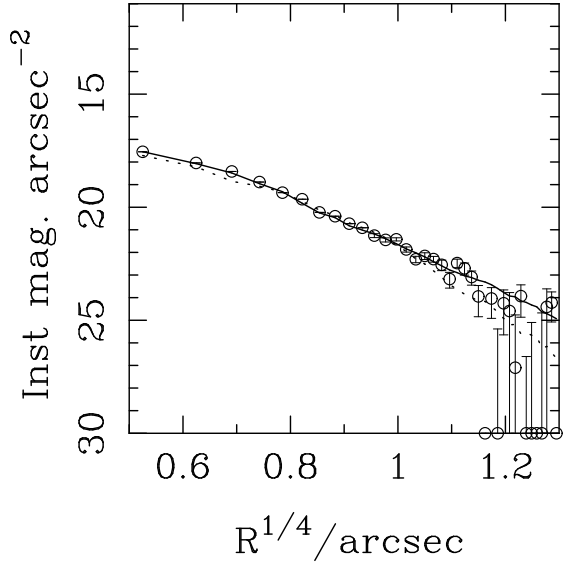}
\caption{\small HST $H$-band profile of 53W069 (circles) compared 
with the best-fitting pure de
Vaucouleurs law after convolution with the NICMOS H-band PSF (solid line)
(Bunker et al. in preparation)}
\end{figure}

\begin{figure}
\vspace{18pc}
\includegraphics{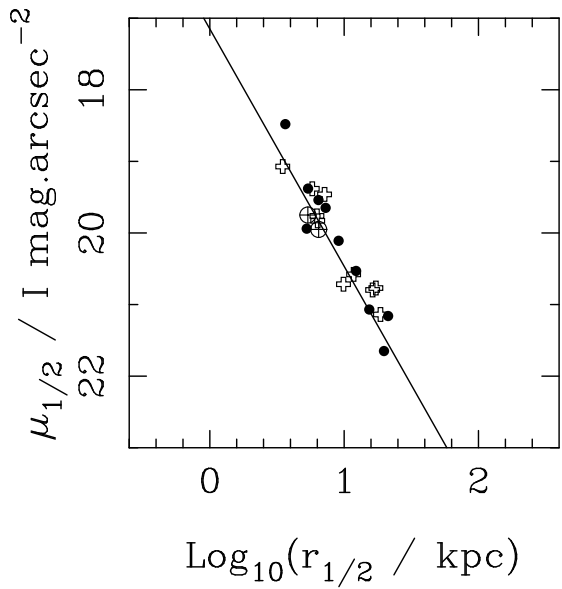}
\caption{\small The Kormendy relation displayed by 3CR radio galaxies at
$z = 0.2$ (solid circles) overlaid with the (indistinguishable) relation
displayed by 3CR galaxies at $z = 0.9$ (crosses) after allowing for 0.6
magnitudes of passive evolution in the $I$-band. The solid line shows the
best fitting Kormendy relation which has a slope of 3.2. The locations 
of 53W091 and 53W069 on this diagram are indicated by the large circle+cross
symbols. They lie on the same Kormendy relation, but with
scalelengths a factor $\simeq 3$ smaller than the mean scalelength
displayed by the more radio-powerful 3CR galaxies.}
\end{figure}

This result, coupled with the rather good agreement between different
models discussed above, suggests that an age limit of $> 3$ Gyr for the
oldest galaxies (and hence the Universe) at $z = 1.5$ should be taken
seriously. Such an age is also consistent with the collapse epoch of
these objects as inferred from the power-spectrum analysis of Peacock et al.
(1998).

\section{Morphology and size}

HST I-band and J-band images of 53W091 and 53W069 indicate that both
galaxies are dominated by a bulge component both above and below the
4000\AA\ break (Waddington et al. in preparation). The most powerful evidence that
morpholically these galaxies are dynamically evolved ellipticals comes
from analysis of the deep NICMOS $H$-band image of 53W091 
obtained by Bunker et al. (in preparation). 
In Figure 5 I show the luminosity
profile derived from this image; a simple de Vaucouleurs law with no
nuclear component provides a significantly better fit to the data than a
disc with a nuclear contribution, and the derived scalelength is $r_e 
= 3.5$ kpc ($\Omega_0 = 1$, $H_0 = 50$km s$^{-1}$ Mpc$^{-1}$).
 
The derived scalelengths and surface brightnesses of both 53W091 and
53W069 are placed in context in Figure 6, which demonstrates that they
lie on the Kormendy relation described by 3CR
radio galaxies (at both high and low redshift; 
McLure \& Dunlop 1999), towards the lower end of 
the scalelength distribution of these more powerful radio galaxies.
As pointed out by McLure \& Dunlop (1999), a reanalysis of the available
HST data on 3CR galaxies provides no evidence for any significant
dynamical evolution of massive ellipticals between $z = 1$; the
location of 53W091 and 53W069 on Figure 6 is not unexpected given their
moderate radio power (see McLure et al. 1999), and 
suggests that the Kormendy relation for massive
ellipticals may already be in place by $z \simeq 1.5$.

\section{Formation epoch inferred from sub-millimetre observations}

\begin{figure}
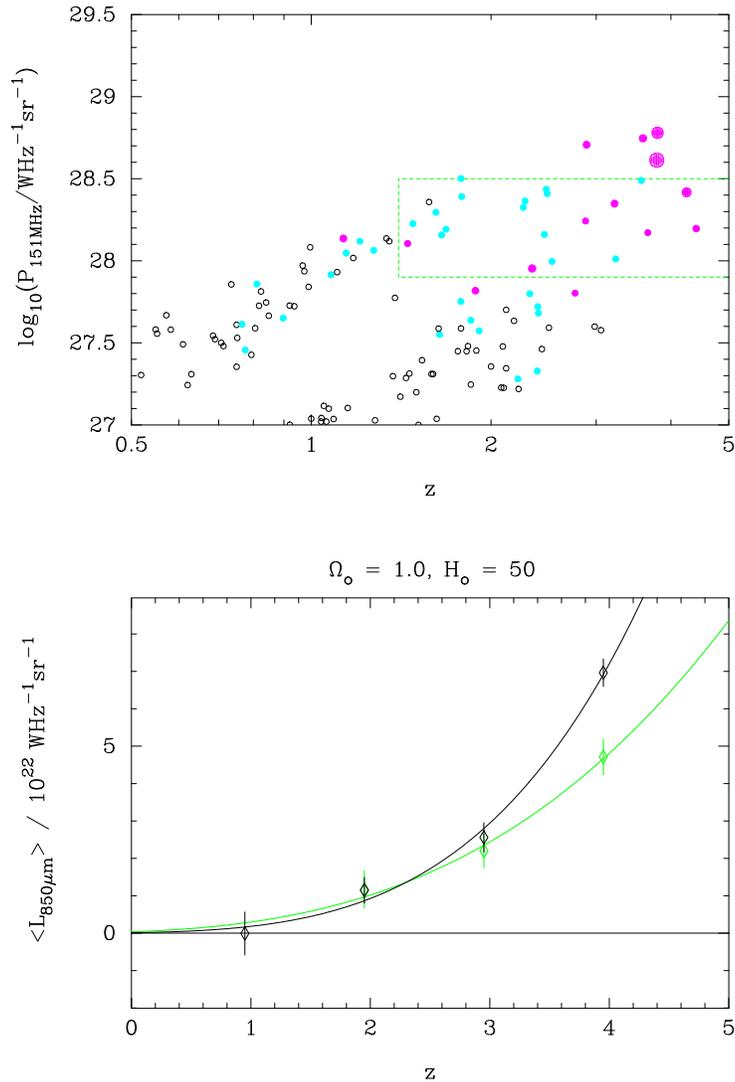

\vspace{35pc}
\includegraphics{dunlopj7a.eps}
\includegraphics{dunlopj7b.eps}
\caption{\small Upper panel: The radio-luminosity redshift plane 
showing the location of the $\simeq 50$ radio galaxies observed with SCUBA by 
Archibald et al. (in preparation). Red symbols of increasing size 
indicate detections of increasing brightness at 
$850 \mu m$. Blue symbols indicate 
non-detections. The green box defines a subsample 
of objects which span the redshift range $z = 1 \rightarrow 4$ at constant 
radio luminosity, allowing separation of the normally correlated effects of 
redshift and luminosity.
Lower panel: Weighted mean $L_{850\mu m}$ for the radio galaxies observed
with SCUBA in redshift bins centred on $z = 1,2,3,4$. The black diamonds
refer to all the radio galaxies observed, while the green diamonds refer
only to the subset of radio galaxies lying within the constant radio
luminosity strip indicated in the upper panel. The lines are fits to the
data of the form $L \propto (1+z)^{\alpha}$ to indicate the strength of
the evolutionary trend. For all galaxies observed $\alpha \simeq 4$, while
for those in the constant luminosity strip $\alpha \simeq 3$. Such
evolution is of very similar strength to that deduced for many other
populations of extragalactic objects out to $z \simeq 2.5$ 
(Dunlop 1998), but here 
appears to continue out to at least $z \simeq 4$. }
\end{figure}

It is interesting to attempt to relate the dynamical and
spectral passivity displayed by 53W091 and 53W069 at $z \simeq 1.5$ to
direct attempts to determine the main epoch of radio galaxy formation
(and possibly all spheroid formation).
There is certainly some evidence that, dynamically, radio galaxies are
different at $z > 3$ than at lower redshifts (van Breugel et al 1998),
but the complications of radio activity have made it
difficult to constrain the major epoch of star-formation activity in
these objects from optical/near-infrared observations.

However, my collaborators and I have recently completed the first major
SCUBA sub-mm survey of dust-enshrouded star-formation activity in radio
galaxies between $z = 1$ and $z = 4$, the results of which indicate that
the main epoch of star-formation activity in these objects lies at $z
> 3$. Previous sub-mm detections of high-redshift radio galaxies have
been at $z \simeq 4$ (Dunlop et al. 1994; Hughes et al. 1997;
Ivision et al. 1998)
but the extreme radio powers of the detected objects made it impossible
to tell whether their inferred large dust and gas masses were primarily
due to cosmic epoch, or instead related to extreme radio power. Now, as
illustrated in Figure 7a, we have achieved sufficient coverage of the
$P-z$ plane to separate these effects, and by considering
a slice at constant radio power, can for the first time 
derive the redshift dependence of
sub-mm emission in powerful radio galaxies. As shown in Figure 7b, the
average sub-mm luminosity (and hence inferred gas mass and star-formation
rate) rises out to at least $z \simeq 4$ with sub-mm luminosity growing
approximately proportional to $(1 + z)^3$. This strongly suggests that
the bulk of star-formation in radio galaxies occurred at redshifts 
$z \simeq 4$ or higher, and that star-formation in radio galaxies,
and arguably ellipticals in general, is close to completion by $z \simeq 3$.

Finally, I note that, in a collaboration led by Rob Ivison, we have 
recently obtained a deep (10-hour) 
SCUBA image of the field centred on
the z = 3.8 radio galaxy 4C41.17. This new image 
reaches an rms sensitivity of $\simeq 1$ mJy at 850$\mu m$ (Ivison et al. in
preparation), and 4C41.17 itself is clearly detected in the centre, 
confirming the early single-element UKT14 detection by Dunlop et al. (1994). 
However, the surprising 
thing about this image is that, despite the fact that 4C41.17 
is one of the most luminous sub-mm sources ever detected, it is {\it
not} the most luminous source in this 2.8-arcmin diameter image.
A more luminous, apparently resolved $\simeq 12$mJy source
lies to the south of the radio galaxy at a projected 
distance of $\simeq 200$ kpc, assuming
it lies at the same redshift (an assumption supported by its sub-mm
colours; Ivison et al. in preparation), and a third moderately bright
source lies to the north. 

Taken together these sources indicate a
considerable excess of dust enshrouded star formation in the vicinity of
this high-redshift radio galaxy compared with that found in blank field
surveys (e.g. Hughes et al. 1998, Eales et al. 1998). Although this is
only a single field, it is strongly suggestive of violent dust-enshrouded
star formation in a high-redshift proto-cluster for which the radio
galaxy has acted as a signpost. Such images provide at
least circumstantial support for the idea that the cosmological 
evolution of dust mass and SF activity plotted in Figure 7b might apply
not just to radio galaxies but to spheroids in general, especially
those born in rich environments (see also Renzini 1999).

\section{Conclusion}

Radio galaxies are obviously a special subset of all ellipticals, and 
very red/passive galaxies such as 53W091 and 53W069 form a special subset
of radio galaxies. However, this does not necessarily mean they are
unrepresentative of ellipticals in general. Rather, it may simply 
mean that it is rare to find elliptical galaxies at $z > 1$ which have 
undergone so little star-formation activity over the $\simeq 3$ Gyrs prior to 
observation. As pointed out by Jimenez et al. (1999c), most realistic 
models of elliptical formation involve sufficient low-level secondary
bursts of star-formation to frequently mask the true optical-infrared 
colours of a dynamically dominant stellar population at $z > 1$ (cf Zepf
1997).

The apparent lack of significant dynamical growth of radio
galaxies since $z \simeq 1$ discussed above (McLure \& Dunlop 1999) may 
also be representative of massive ellipticals in general.
Certainly there is now a growing body of observational evidence from
infrared studies that
cluster ellipticals (de Propris et al. 1999) and field ellipticals
(Dunlop et al. 1999) are essentially all in place by $z \simeq 1$
(cf Kauffman \& Charlot 1998). 

This lack of dynamical action since $z \simeq 1$, and the ages of the
oldest ellipticals at $ z > 1$, both favour a universe with
$\Omega_{matter} < 1$.

\acknowledgments
This paper draws on recent results from a number of programmes,
and I gratefully acknowledge the contributions of my collaborators Hy
Spinrad, John Peacock, Raul Jimenez, Arjun Dey, Daniel Stern, Rogier
Windhorst, Ian Waddington, Louisa Nolan, Ross McLure, Andy Bunker, David
Hughes, Elese Archibald, Rob Ivison, Steve Rawlings and Steve Eales.

\end{document}